# Direct observation of state-filling at hybrid tin oxide/organic interfaces

Ulrich Hörmann[1*], Stefan Zeiske[1], Soohyung Park[2,3], Thorsten Schultz[2], Sebastian Kickhöfel[2], Ullrich Scherf[4], Sylke Blumstengel[2], Norbert Koch[2,3], and Dieter Neher[1*]

[1]Institute of Physics and Astronomy, University of Potsdam, Potsdam, Germany

[2]Institute of Physics, Humboldt University of Berlin, Berlin, Germany

[3]Helmholtz-Zentrum für Materialien und Energie GmbH, Bereich Solarenergieforschung, Berlin, Germany

[4]Macromolecular Chemistry Group, University of Wuppertal, Wuppertal, Germany

* Corresponding authors: ulrich.hoermann@uni-potsdam.de, neher@uni-potsdam.de

**Abstract**

Electroluminescence (EL) spectra from hybrid charge transfer excitons at metal oxide/organic type-II heterojunctions exhibit pronounced bias-induced spectral shifts. The reasons for this phenomenon have been discussed controversially and arguments for both electric field-induced effects as well as filling of trap states at the oxide surface have been put forward. Here, we combine the results from EL and photovoltaic measurements to eliminate the disguising effects of the series resistance. For $SnO_x$ combined with the conjugated polymer MeLPPP, we find a one-to-one correspondence between the blueshift of the EL peak and the increase of the quasi-Fermi level splitting at the hybrid heterojunction, which we unambiguously assign to state filling. Our data is resembled best by a model considering the combination an exponential density of states with a doped semiconductor.

In analogy to the donor/acceptor interface in organic solar cells, hybrid inorganic/organic interfaces may form staggered type-II heterojunctions that are able to generate charge carriers from optically excited excitons.[1–6] Much like for all-organic systems, the open circuit voltage ($V_{oc}$) generated by such hybrid junctions is closely related to the energy of the charge transfer (CT) state formed at the interface between the inorganic semiconductor and the organic compound.[7] Despite this relation, it has been reported by different groups that the spectral position of the electroluminescence (EL) signal originating from these hybrid CT states (HCTS) depends on the driving conditions. Some groups, including ourselves, have found that the observed energy shift is primarily induced by the internal electric field in the organic component.[8–10] Others have argued that the shift is observed because one species of charge carriers forming the HCTS is bound by traps.[11] In first approximation, the energy of the CT state is given by the energy difference between the electron and hole, constituting the CT exciton. Increased state filling will then also increase the energy of the CT composite state. This situation is schematically illustrated in Fig. 1 (a) for the particular case of an exponential trap distri-bution for electrons in the conduction band (CB) of the inorganic semiconductor and a Gaussian density of states (DOS) (exp-Gauss) for holes in the highest occupied molecular orbital (HOMO) of the organic material. Here, the exponential and Gaussian DOS are given by:

$$g_{\exp}(E) = \frac{N_C}{k_B T_0} \exp\left(\frac{E}{k_B T_0}\right) \quad (1)$$

and

$$g_{\text{Gauss}}(E) = \frac{N_V}{\sigma\sqrt{2\pi}} \exp\left(-\frac{E^2}{2\sigma^2}\right), \quad (2)$$

respectively, where $T_0$ denotes the trap temperature, $N_C$ and $N_V$ are the effective densities of states, and $\sigma$ specifies the width of the Gaussian.

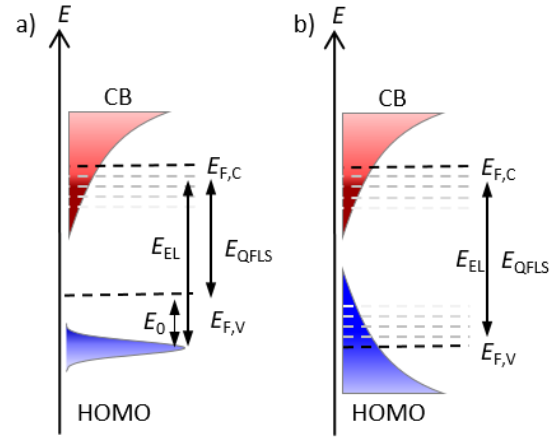

Figure 1 Cartoon of the model densities of states. For the combination of exponential (here electrons) and Gaussian DOS (here holes) (a), the energy of the CT state is approximately given by the difference between the HOMO energy and the electron quasi-Fermi level. In the case of two exponentials (b) the CT state energy is directly determined by the QFLS.

Approximating the occupation function in the exponential DOS with a step function [i.e. all states below (above) the Fermi level are occupied (empty)] and assuming Boltzmann population of the Gaussian DOS, we obtain the electron density in the exponential tail of the conduction band as[12]

$$n_{\text{exp}} = N_C \exp\left(-\frac{E_C - E_{F,C}}{k_B T_0}\right), \quad (3)$$

and the hole density in the Gaussian DOS as

$$p_{\text{Gauss}} = N_V \exp\left(-\frac{E_{F,V} - E_V - \frac{\sigma^2}{2k_B T}}{k_B T}\right). \quad (4)$$

Here, $E_C$ denotes the reference energy of the CB and $E_V$ is the energy of the maximum of the Gaussian DOS. $E_{F,C}$ and $E_{F,V}$ are the quasi-Fermi levels of electrons in the conduction and valance bands, respectively. The maxima of the charge carrier populations are then located at the Fermi-level for the exponential DOS and at $E_\infty$, i.e., $\frac{\sigma^2}{k_B T}$ above $E_V$, for the Gaussian profile.[12] Neglecting the exciton binding energy, the peak position of the EL signal, resulting from the recombination of electrons and holes, can then be approximated by the energy separation of their respective populations. With $E_G = E_C - E_V$ we obtain

$$E_{\text{EL}} = E_{F,C} - \left(E_V + \frac{\sigma^2}{k_B T}\right) = E_G - \frac{\sigma^2}{k_B T} - (E_C - E_{F,C}) = E_G - \frac{\sigma^2}{k_B T} + k_B T_0 \ln\left(\frac{n_{\text{exp}}}{N_C}\right). \quad (5)$$

The quasi-Fermi-level splitting ($E_{\text{QFLS}} = E_{F,C} - E_{F,V}$) can then be expressed as

$$E_{\text{QFLS}} = E_G - (E_C - E_{F,C}) - (E_{F,V} - E_V) = E_G + k_B T_0 \ln\left(\frac{n_{\text{exp}}}{N_C}\right) - \frac{\sigma^2}{2k_B T} + k_B T \ln\left(\frac{p_{\text{Gauss}}}{N_V}\right). \quad (6)$$

Combining (5) and (6) we can finally write the EL peak position in terms of the QFLS:

$$\begin{aligned} E_{\text{EL}} &= E_{F,C} - E_\infty \\ &= E_{\text{QFLS}} \underbrace{- \frac{\sigma^2}{2k_B T} - k_B T \ln\left(\frac{p_{\text{Gauss}}}{N_V}\right)}_{E_0 = -(E_\infty - E_{F,V})}. \quad (7) \end{aligned}$$

Importantly, Eq. (7) shows that the EL peak position is not an explicit function of the QFLS. This is because in the limit considered here, the energy $E_\infty$ where the majority of holes is located in the Gaussian does not depend on the hole quasi-Fermi-level $E_{F,V}$. This is accounted for by the additional terms in Eq. (7).

Assuming charge neutrality, i.e., $n_{\text{exp}} = p_{\text{Gauss}}$, Eq. (7) can be re-written as

$$E_{\text{EL}} = \frac{T_0}{T + T_0} E_{\text{QFLS}} + \frac{T}{T + T_0} E_G - \left(1 + \frac{T}{T + T_0}\right) \frac{\sigma^2}{2k_B T} + \frac{k_B T T_0}{T + T_0} \ln\left(\frac{N_V}{N_C}\right), \quad (8)$$

where the last term vanishes if the effective densities of states are the same for electrons and holes. From Eq. (8) we can now see that, if trap states of only one carrier species [cf. Fig. 1 (a)] are involved in the recombination process, and if charge neutrality is fulfilled, the EL peak position will scale linearly with the QFLS with the slope solely determined by the trap and the device temperatures.

If, on the other hand, the semiconductor is p-doped at the interface between donor and acceptor, the excess hole concentration $\Delta p$, generated by light or injected through contacts, can be neglected. The total hole density $p = p_0 + \Delta p \approx p_0$ remains practically constant at the dark hole density $p_0$. Vice versa, the dark carrier density of electrons is negligible and the total electron density $n$ is approximately given by the excess electron density $\Delta n$. From Eqs. (6) and (7) it is evident that under such circumstances the hole quasi-Fermi-level is a material specific parameter that does not depend on the device operation conditions apart from the device temperature. In that case the EL peak position does indeed scale directly with the QFLS with unity slope:

$$E_{\text{EL}} = E_{\text{QFLS}} - \frac{\sigma^2}{2k_B T} - k_B T \ln\left(\frac{p_0}{N_V}\right). \quad (9)$$

It is instructive to compare this result to the situation, where both electrons and holes reside in exponential DOS [exp-exp, cf. Fig. 1 (b)]. With the step-occupation

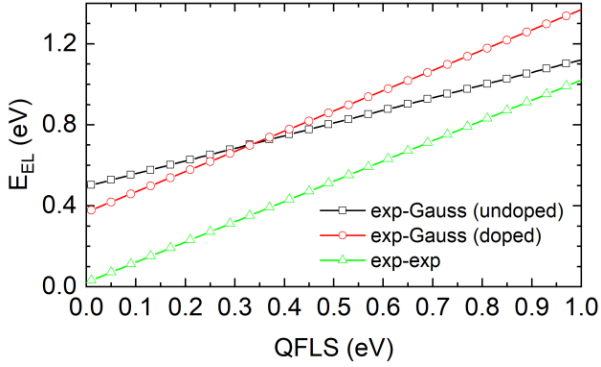

Figure 2 Calculated correlation of the EL peak position with the QFLS for the configurations described above. In the undoped exp-Gauss case, the slope is less than one [cf. Eq. (8)]. The doped exp-Gauss case [Eq. (9)] shows unity slope but a finite intercept. For the exp-exp case [Eq. (10)] we get unity slope and an intercept very close to zero. The used parameters are $T = 300$ K, $T_0 = 500$ K, $\sigma = 0.05$ eV, $E_g = 1.4$ eV, $N_V = 10^{20}$ cm$^{-3}$, $N_C = 10^{19}$ cm$^{-3}$, $p_0 = 10^{13}$ cm$^{-3}$.

assumption and the hole density $p_{\exp}$ expressed in analog to Eq. (3) we arrive at

$$E_{\text{EL}} = E_{\text{QFLS}} = E_G + k_B T_{0,C} \ln\left(\frac{n_{\exp}}{N_C}\right) + k_B T_{0,V} \ln\left(\frac{p_{\exp}}{N_V}\right). \quad (10a)$$

If the trap distributions of electrons and holes as well as their carrier densities are equal, Eq. (10a) simplifies to:

$$E_{\text{EL}} = E_{\text{QFLS}} = E_G + 2k_B T_0 \ln\left(\frac{n_{\exp}}{N_C}\right). \quad (10b)$$

The EL-QFLS relationships [Eqs. (8), (9) and (10)] are compared graphically in Fig. 2. We note that for both exponential with doped Gaussian and two exponentials, the EL peak position scales directly with the QFLS. However, in the second case, the energy offset at zero QFLS is negligible. This is in congruence with the fact that in the presence of exponential tails the concept of an energy gap is not valid any more.

From an experimental perspective, it is therefore important to take a closer look at the temperature dependence of the QFLS. The linear extrapolation of the QFLS (measured as the open circuit voltage) to zero Kelvin is often used to infer the energy of the CT energy gap.[9,13–16] It is, however, apparent from Eq. (10) that in the exp-exp case the QFLS is temperature independent, unless the recombination process and, hence, the carrier densities themselves depend on temperature. In the exp-Gauss case (doped or not) [Eq. (6)], the dependence on temperature will initially be linear, but as temperature decreases the influence of the broadened Gaussian DOS will become more and more dominant, as shown in Fig. 3. Note that the predicted QFLS roll-off at low temperatures is intrinsic to the considered DOS combination and not related to contact or transport properties. Furthermore, regardless of doping, the carrier density in the exponential tail will add a temperature independent offset to the QFLS. This implies that linear extrapolation of the high temperature range of the QFLS measured at different light intensities does not coincide in a single energy at 0 K (illustrated by the solid lines in Fig. 3); this is, however, necessary for meaningful extraction of the photovoltaic gap and often fulfilled for all-organic devices.

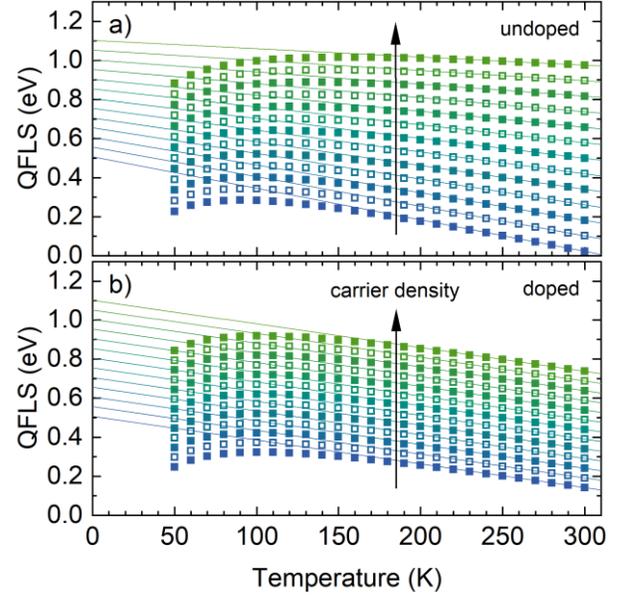

Figure 3 Temperature and carrier density (i.e. light intensity) dependent QFLS without (a) and with (b) doping, calculated from Eq. 6. The solid lines are the linear extrapolation of the high temperature regime, like it would be carried out in order to determine the photovoltaic gap.

For hybrid type-II heterojunctions formed by a transparent metal oxide and an organic semiconductor, it has been disputed if state filling does play a role for the experimentally observed EL peak shift.[8–11] Recent publications put forward convincing arguments against this hypothesis.[9,10] However, we could now identify a real world system that shows the behavior predicted by the above state-filling models: The planar heterojunction between SolGel deposited $SnO_x$ and ladder-type poly-(para phenylene) (MeLPPP)[17,18] corresponds particularly well to the doped exp-Gauss case. Fig. 4(a) shows the CT EL signal of such a device (ITO/ $SnO_x$/ MeLPPP/ $MoO_3$/ Al) recorded at stepwise increased current densities [solid circles in Fig. 4 (b)]. The blueshift of the EL peak with increasing current

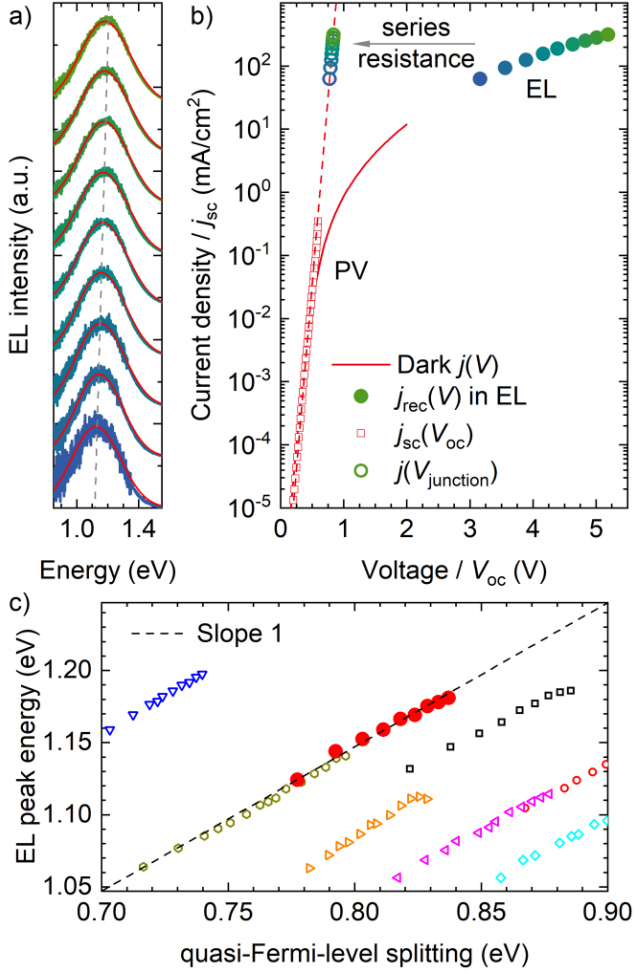

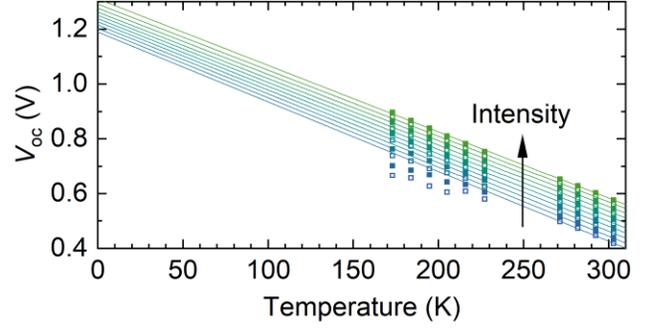

Figure 5 Open circuit voltage of an SnO$_x$/MeLPPP device measured as a function of temperature and light intensity (405nm LED). The linear extrapolation (solid lines) at different light intensities does clearly not coincide at 0 K.

Here $n_{id}$ is the diode ideality factor and $j_0$ is the dark saturation current.

At $V_{oc}$ the recombination current is exactly cancelled by the generation current and the QFLS is directly measured. Under identical illumination conditions, the generation current is then approximated by the measured $j_{sc}$. Using Eq. (11) we can then estimate the QFLS associated with a given recombination current under EL driving conditions, essentially projecting the current density onto the extrapolated $j_{sc}(V_{oc})$ curve [dashed line in Fig. 4 (b)].

The correlation between the EL peak energy and the QFLS extracted by this method is shown in Fig. 4 (c) for different SnO$_x$/MeLPPP samples. The solid circles correspond to the sample in Fig. 4 (a) and (b). In all cases the peak position clearly follows the QFLS with unity slope as indicated by the dashed line, which serves as a guide to the eye. Despite the identical slope across different samples, the intercept varies between 0.42 eV and 0.26 eV. This behavior matches with the doped exp-Gauss model. From Eq. (9) it is apparent that the intercept depends on the doping concentration $p_0$ and the width of the Gaussian. Both parameters are likely to be subject to large sample to sample variation.

The temperature and intensity dependence of $V_{oc}$ of an exemplary SnO$_x$/MeLPPP device is shown in Fig. 5. The qualitative similarity with Fig. 3 (b) is striking and the parallel shift of the measurement curves by the increasing light intensity is a clear sign of the presence of an exponential trap distribution and in accordance with the findings from Fig. 4 (c).

By comparing the experimentally determined relationship between EL peak position and QFLS, and the temperature dependent $V_{oc}$ at different light intensities (Fig. 4 (c) and 5) with the predictions from density is substantial and comparable to UV-protected ZnO/organic devices that could not be described by our field effect model.[10]

Figure 4 EL spectra (a) of an SnO$_x$/MeLPPP hybrid device, driven at different current densities [solid circles in (b)]. (b) Reconstruction of the QFLS (open circles) associated with the recombination current is achieved by projection of the current density onto the extrapolation of the intensity dependent $j_{sc}(V_{oc})$ measurement, which removes the influence of the series resistance, visbible in the dark $j(V)$ curve. The EL peak position determined by Gaussian fits [red lines in (a)] is plotted against the QFLS in (c). The solid circles belong to the sample in (a) and (b), the open symbols are the data derived from a series of similar samples with different organic and inorganic semiconductor thicknesses.

In order to determine the QFLS associated with the peak positions (determined by Gaussian fits), we reconstruct the series resistance-free $j(V)$ curve by plotting the short circuit current ($j_{sc}$) measured at different light intensities against the corresponding $V_{oc}$ value [open rectangles in Fig. 4 (b)]. Here, we take advantage of the fact that the recombination current $j_{rec}$ across the junction is a sole function of the QFLS[19]:

$$E_{QFLS} = eV_{junction} = n_{id}k_B T \ln\left(\frac{j_{rec}}{j_0} + 1\right). \quad (11)$$

the considered DOS constellations (Fig. 2 and 3), we conclude that the observed shift of the EL peak with increasing driving current is indeed caused by state-filling. Furthermore, it appears that the combination of an exponential trap distribution with a doped Gaussian resembles our data best. We would like to stress that even though the models were derived assuming trapped electrons recombining with free holes, the results cannot be distinguished from the opposite case where an n-doped semiconductor is combined with an undoped material exhibiting a broad trap distribution above the valance band onset (or HOMO). We can therefore not conclusively identify if the traps are located on the $SnO_x$ or the polymer side of the junction. However, photoelectron spectroscopy results (see SI) indicate that our $SnO_x$ is n-doped. It seems, hence, possible that a significant amount of traps is present in the MeLPPP, possibly even caused by the metal oxide through contact-induced DOS broadening.[20,21] Regardless of this uncertainty, our data obtained by reconstruction of the QFLS associated with the EL driving current clearly evidence that current dependent state-filling can indeed be the cause of a spectral shift of the CT feature. Together with earlier reports that identify the peak shift as an electric field effect, this illustrates that a general assignment of the origin cannot easily be made and precise knowledge of the exact hybrid system is necessary.

This work was funded by the German Research Foundation (DFG) within the collaborative research center 951 "HIOS".

$SnO_X$ films were prepared by spin coating a Sn(II)-Cl precursor solution in ethanol onto ITO covered glass slides. MeLPPP (provided by U. Scherf, Uni Wuppertal) was spincast from solution in chlorobenzene. $MoO_3$ and aluminum were used as the hole injecting contact and deposited onto the MeLPPP film by thermal evaporation in vacuum. The EL setup is described elsewhere.[7] $j(V)$ curves and $j_{sc}(V_{oc})$ data were recorded with a Keithley 2400 sourcemeter. Intensity dependent device characterization was performed by using a 405 nm LED at different driving currents. A closed cycle He cryostat was used for temperature dependent measurements.

# Supporting Information

**Photoelectron and inverse photoelectron spectroscopy data**

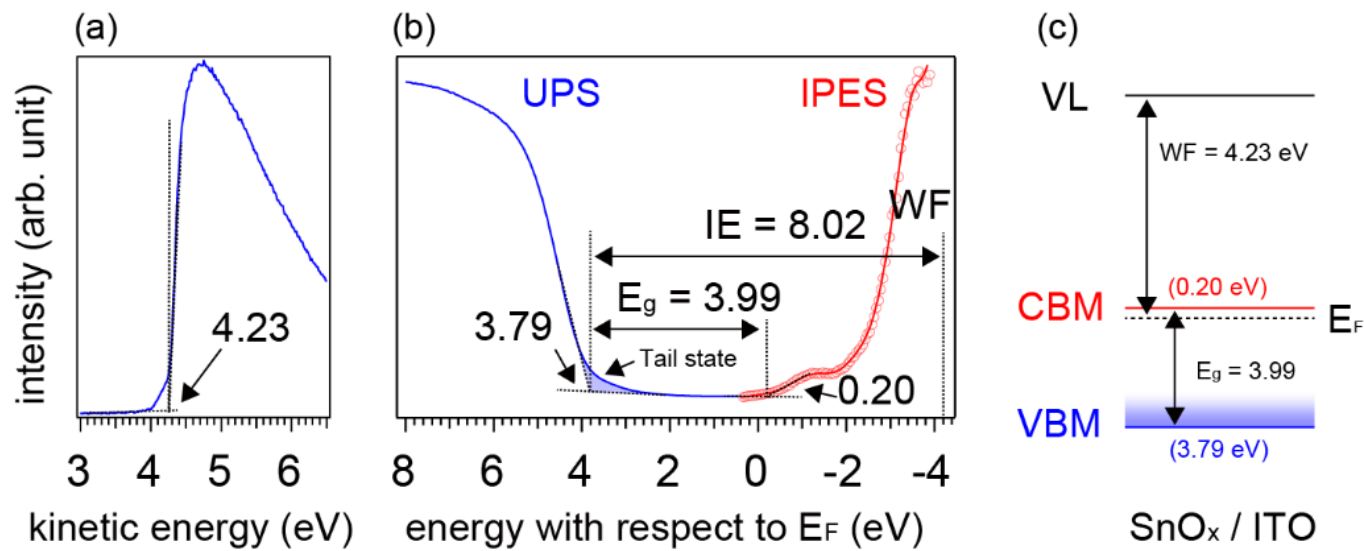

Figure 6 UPS and IPES data of our SolGel-processed $SnO_x$ films. The SECO is shown in (a), The valence and conduction band energies w.r.t. the Fermi-level are shown in (b). The film is clearly n-doped and shows significant tail states in the valence band, which might lead to contact induced broadening of the polymer DOS when the heterojunction is formed. He $I_\beta$ satellite emission features were removed from measured valence band spectra. Likewise, IPES spectra were deconvoluted as described in a previous report. [DOI:10.1088/2053-1583/aaa4ca]. An overview of the relevant energies is shown in the scheme in (c).